\documentclass[%
 reprint,
nofootinbib,
 amsmath,amssymb,
 aps,
]{revtex4-1}

\usepackage{graphicx}
\usepackage{dcolumn}
\usepackage{bm}
\usepackage[utf8]{inputenc}
\usepackage{amssymb}
\usepackage{tensor}
\usepackage{xspace} 
\usepackage[pdftex]{hyperref}
\usepackage{hyperref}
\usepackage{graphicx}
\usepackage[caption=false]{subfig}
\usepackage{breakcites}
\usepackage{ragged2e}
\usepackage{cleveref}

\usepackage{setspace}

\hypersetup{
colorlinks,linkcolor=blue,citecolor=blue,urlcolor=blue
}

\usepackage{natbib}
\usepackage{tablefootnote}
\usepackage{footnote}
\usepackage{amsmath}
\usepackage{amssymb}
\usepackage[T1]{fontenc}

\begin{document}

\preprint{APS}


\title{The amplification of cosmological magnetic fields in Extended $f(T,B)$ Teleparallel Gravity}

\author{Salvatore Capozziello$^{a,b,c}$, Amodio Carleo$^{d,e}$,  Gaetano Lambiase$^{d,e}$}
\email{capozziello@na.infn.it}
\affiliation{$^a$Dipartimento di Fisica "E. Pancini", Universit\'a degli Studi di Napoli Federico II", Via Cinthia, I-80126, Napoli, Italy}
\affiliation{$^{b}$Istituto Nazionale di Fisica Nucleare (INFN), sez. di Napoli, Via Cinthia 9, I-80126 Napoli, Italy}
\affiliation{$^{c}$Scuola Superiore Meridionale, Largo S. Marcellino, I-80138, Napoli, Italy}

\email{acarleo@unisa.it}
\affiliation{$^{d}$Dipartimento di Fisica "E. R. Caianiello", Universit\'a degli Studi di  Salerno,  Via Giovanni Paolo II, 132, I-84084, Fisciano (SA), Italy}
\email{lambiase@sa.infn.it}
\affiliation{$^{e}$Istituto Nazionale di Fisica Nucleare (INFN), gruppo collegato di Salerno,  Italy}




\date{\today}

\begin{abstract}
Observations indicate that intergalactic magnetic fields have amplitudes of the order of $\sim 10^{-6}$ G and are uniform on scales of $\sim 10$ kpc. Despite their wide presence in the Universe, their origin remains an open issue. Even by invoking a dynamo mechanism or a compression effect for magnetic field amplification, the existence of  seed fields before galaxy formation is still problematic. General Relativity predicts an adiabatic decrease of the magnetic field evolving as $|\mathbf{B}|\propto 1/a^{2}$, where $a$ is the scale factor of the Universe. It results in very small primordial fields, unless the conformal symmetry of the electromagnetic sector is broken. In this paper, we study the possibility that a natural mechanism for the  amplification of primordial magnetic field can be related to extended teleparallel gravity  $f(T, B)$ models, where $T$ is the torsion scalar, and $B$ the boundary term. In particular,  we consider a non-minimal coupling with gravity in view to break conformal symmetry in a teleparallel   background, investigating, in particular, the role of  boundary term $B$, which can be consider as a further scalar field. We find that, after solving exactly the $f(T,B)$ field equations both in inflation and reheating eras, a non-adiabatic behavior of the  magnetic field is always possible,  and  a strong  amplification appears  in the reheating epoch. We also compute the ratio $r=\rho_{B}/ \rho_{\gamma}$ between the magnetic energy density and the cosmic microwave energy density during inflation,  in order to explain the present value $r\simeq 1$, showing that, in the slow-roll approximation, power-law teleparallel theories with $B^{n}$  have effects indistinguishable from metric theories $R^{n}$ where $R$ is the Ricci curvature scalar..  
\end{abstract}

\maketitle


\section{Introduction}\label{sec1}
The presence of magnetic fields in any structure of our Universe is a  consolidated issue confirmed by several fine observational data. Our Galaxy and other spiral galaxies are endowed with coherent large scale magnetic fields with typical length $\ge 10$ kpc and  strength of $\sim 3 \times 10^{-6}$ G. They play  important roles in a multitude of astrophysical phenomena, such as the confinement of cosmic rays, the transfer of angular momentum away from protostellar clouds (allowing their collapse in stars), the genesis of gamma ray-bursts (GRBs) and, recently, the  extraction of energy from BHs \cite{Comisso,Carleo:2022qlv,Khodadi_2022,Wei_2022}. While local strong magnetic fields (up to $10^{14}$ G) come out from stars or compact objects (like neutron stars), galactic and intergalactic magnetic fields still have no explanation, being one of the long standing problem of astrophysics and cosmology. While in the case of aged galaxies, one could explain them by invoking a dynamo \cite{Widrow_2002} or a compression mechanism \cite{Turner:1987bw}, the presence of fields also in protogalaxy and mostly in the intergalactic medium (IGM), suggests their cosmological rather than local origin. This eventually means to search for some unknown physical process for generating such large scale fields. One possibility is that they are relics from the early Universe, with a subsequent amplification in a pre-galactic era through dynamo or compression effects. In the first case, a galactic dynamo for the entire age of the galaxy could
have amplified a primordial magnetic field (PMF) by a factor of $10^{13}$, thus requiring a primordial field of $10^{-19}$ G; in the second case, the strength of the seed field required to explain today galactic values is much greater, $10^{-9} G$ \cite{Turner:1987bw}, therefore resulting less efficient. Such a primordial magnetic field, also called seed field, however, have not a clear origin and its evolution and signatures are still a matter of debate today.  The size of the initial magnetic seed is also an issue, since it should not be smaller that $100$ pc after the collapse of the protogalaxy, which implies a comoving scale of approximately $10$ Kpc before the collapse. Seeds generated after inflation, during the radiation era, for example, are typically too small in size because their coherence length can never exceed that of the causal horizon at the time of magnetogenesis. Inflation remains the only mechanism capable of  producing super-horizon correlations, so, in principle,  it could easily generate primordial fields of the required length. 

Specifically, General Relativity (GR) predicts an adiabatic decay for the PMF: since the Universe is believed to have been a good conductor for much of its  post-inflationary epochs, any  cosmological
magnetic field, from the end of inflation onwards,  will preserve the  flux, i.e. $a^2 B \sim const $, and then $B \sim 1/a^2$, or, in terms of the magnetic energy density, $\rho_{B} = |\mathbf{B}|^{2}/(8 \pi ) \propto 1/a^4$, where $a$ is the scale factor of the (flat) Friedman-Robertson-Walker (FRW) metric. Since the scale factor tends to infinity during inflation, this type of decay involves too many weak or practically absent magnetic fields at the end of the inflation period. This scaling is the same for every cosmic energy density present in the Universe. In particular, the Universe is filled with a cosmic
microwave background radiation (CMB), a relic of the hot big bang, with a thermal spectrum at the (current) temperature of $T = 2.725$ K. The energy density of this radiation, $\rho_{\gamma} = \pi^{2} T^{4}/ 25$, which formed
a dominant component of the energy density of the
early Universe, corresponded to the energy of the void and was almost constant during inflation. Immediately afterwards, it began to dilute as the Universe expands as $\rho_{\gamma} \sim 1/a^4$ (the extra factor $1/a$ w.r.t to matter, which decays as $\sim 1/a^{3}$, comes from energy redshift); therefore, the ratio $r \doteq \rho_{B}/ \rho_{\gamma}$   remained constant until today, with a current value of $r \approx 1$. It is then
standard practice to characterize the primordial field
with either this ratio $r$, or the present day value $\mathbf{B}_{0}$ as
a function of its coherence scale $L$. Precisely, a present day
magnetic field strength of $3.2 $ $\mu G$ has an energy density equal
to the present day CMB energy density, i.e. $r = 1$. In order to explain this value, one needs a  pregalactic seed field with a ratio $r \simeq 10^{-34}$ if dynamo amplification is assumed, and $r \simeq 10^{-8}$ if compression occurred in the collapse of protogalactic cloud.   
 
This problem has been studied not only in the context of GR but also e.g. in the Poincare gauge theory \cite{Kothari:2018aem}, string theory \cite{Gasperini:1995dh}, Gauss-Bonnet gravity \cite{Atmjeet:2013yta} and   gravity theories with torsion \cite{Kothari:2018aem}. An adiabatic decay-law on all length  scales, moreover, implicitly assumes the existence of electric currents with super-horizon
correlations, thus violating causality \cite{Tsagas}. One way to overcome these critical points is to assume that the flux is not conserved and therefore that the electromagnetic sector is no longer conformally invariant, i.e. $\int d^{4}x F^{\mu \nu} F_{\mu \nu} \not= \int d^{4}\Tilde{x} \Tilde{F}^{\mu \nu} \Tilde{F}_{\mu \nu}$, where $F^{\mu \nu}$ is written in terms of the new metric $\Tilde{g}_{\mu \nu} = \Omega^{-2} g_{\mu \nu}$, with $\Omega$ the conformal factor.

This would imply a non-adiabatic decay of the  primordial magnetic field $\mathbf{B}$, ensuring its survival even after inflation, in the form of  current large-scale fields.  

On the other hand, post-inflationary scenarios consider PMFs created after inflation via either cosmological phase transitions or during the recombination era \cite{Subramanian:2015lua}, even if,  in this case, the role of helicity is fundamental  to transfer energy from small to large scales \cite{Durrer:2013pga}. For other mechanisms, aimed at explaining the origin and the amplification, see  \cite{Giovannini:2004yn} for a review. 

Moreover, it has also  been pointed out that  a FRW  negative curvature $(K=-1)$ allows a super-adiabatic decay, i.e. in this case the magnetic field would have had a relative amplification w.r.t. the radiation.   

In order to have a large scale field, i.e. causally disconnected, the PMF must have crossed outside the Hubble radius $\sim H^{-1}$ during de Sitter phase, i.e. with a typical wavelength $\lambda_{phys} \gg H^{-1}$ or $k \eta \ll 1$, where $\eta$ is the conformal time during inflation, leading  to a static, large scale $\mathbf{B}$. These long-wavelength modes, in particular, may have been generated by quantum fluctuations which grew   during inflation and  the reheating eras,  in a similar way to those which led to  density fluctuations  and thus to the large scale structure. Their effect on gravitational waves (GWs) and on the CMB has been studied in several papers, see, for example
\cite{GW-2021al,Kunze:2013kza,Addazi:2022ukh}.

In this debate, a particular role has been recently assumed by Teleparallel Gravity (TG) and its extensions which seem to fix several cosmological issues ranging from inflation to dark energy, from primordial Big Bang nucleosynthesis to cosmological perturbations, up to the $H_0$ tension
\cite{Ferraro:2006jd,Dent:2010nbw,Chen:2010va,Cai:2015emx,Benetti:2020hxp,Capozziello:2017bxm, Escamilla-Rivera:2019ulu}. The approach
starts from the so called Teleparallel Equivalent General Relativity (TEGR) \cite{Maluf:2013gaa} which is an alternative formulation of GR, firstly conceived by Einstein himself, where dynamical variables are tetrads and dynamics is given by the torsion instead of curvature in a teleparallel affine formulation of gravity. 

TG and its extensions could have a prominent role in generating and amplifying primordial magnetic fields.

In this paper, we want to investigate whether an alternative theory of gravity, the $f(T,B)$ theory where $T$ is the torsion scalar and $B$ a boundary term, with a non-minimal coupling between matter and gravity, can generate primordial magnetic fields with a non-adiabatic behaviour, exploiting the breaking of conformal symmetry which naturally arises from such non-minimal couplings.  These couplings are well motivated since according to the Quantum Electrodynamics (QED) formulated on curved space-times, one-loop vacuum-polarization effects \cite{Drummond:1979pp} can lead to non-minimal gravitational couplings between the curvature and the electromagnetic field.  

The electromagnetic sector $F_{\mu \nu}F^{\mu \nu}$ is assumed coupled to the curvature Ricci scalar $R$ and Riemann tensor $R^{\mu \nu \rho \sigma}$ curvatures in \cite{Turner:1987bw,Lambiase:2004zb,Lambiase:2008zz,deAndrade:2013fga}. On the other hand,  curvature power-law models $R^{n}$ are considered in \cite{Garretson:1992vt,Mazzitelli:1995mp,Lambiase:2008zz,Bertolami:2022hjk}, while  torsion power-law models $T^{n}$ in \cite{Bamba:2013rra}.  Specifically, we are going to adopt here a TG background where $T$  and $B$ are related to curvature scalar as  $R = -T + B$.  In this context, it results interesting to explore the role of the boundary term $B$ which link the metric picture to the teleparallel one. In fact, $f(R)$ gravity is not equivalent to $f(T)$ gravity \cite{Bamba:2013ooa} since the latter is not invariant for local Lorentz transformations and has second order field equations (see \cite{Cai:2015emx} for a review). To restore fourth-order field equations, one has to introduce the boundary term $B$ which depends on the derivatives of the torsion vector $T^{\mu}$, obtaining the $f(T,B)$ theory which, as a special case, reduce to  $f(R)$ \cite{Bahamonde:2016grb,Capozziello:2019msc}. This is the only framework to investigate the role of $B$ separated from $R$ and $T$. Finally, as noted in \cite{Kranas:2018jdc}, a torsion dominated early universe could go trough a phase of accelerated expansion without the need of a cosmological constant or dark energy component, thus constituting an interesting theoretical framework to be studied. 
In this paper, we are going to investigate the role of $T$, $B$, and eventual non-minimal couplings to matter field to enhance PMFs to make them compatible with observations.

The layout of the paper is as follows: in Sec. \ref{sec2} we review  the teleparallel $f(T,B)$ gravity theory. Sec. \ref{sec3} is devoted to computing and solving the  cosmological equations in a spatially flat FRW metric, distinguishing between inflationary and reheating eras. In the Sec. \ref{sec4} we consider a non-minimal coupling gravity-photon and we obtain a differential equation for the magnetic field, which we are going to solve for both inflation and reheating epochs. A different approach to evaluate the amplification effect during inflation is treated in Sec. \ref{sec5}, while discussion and conclusions are drawn in the Sec. \ref{sec6}.   
 
In this work, we adopt  natural units  $\hbar=c=1$, and we define the Planck
mass  be $M^{2}_{pl} = {8 \pi G}$. The metric signature  $(+, -, -, -)$ is also adopted. The Greek indices are coordinate ones contracted with the metric tensor $g_{\mu \nu}$, the
Latin indices $a,b,c$ are tetrad ones  contracted with the Kronecker delta. The Latin indices $i,j$ indicate the spatial coordinates. The overdots and overprimes, as usual, denote derivatives with respect to the cosmic time $t$ (measured by a comoving observer) and the conformal time $\eta$, respectively, unless otherwise stated.

\section{Field equations in $f(T,B)$ gravity}
\label{sec2}

Extensions or modifications of GR are becoming an ideal arena on which longstanding cosmological problems can be  solved or incorporated into the theory itself from a gravitational viewpoint \cite{Capozziello:2011et}.  TG   differs from  metric  theories for several features, among  them  a non-zero torsion, which arises from a non-symmetric connection\cite{Capozziello:2022zzh}. A property of TG is that  the Lovelock theorem is weakened \cite{Gonzalez:2015sha}, allowing for additional theories beyond TEGR that continue to produce second order field equations \cite{Cai:2015emx}. The teleparallel boundary term $B$ (see later) embodies the fourth order contributions to the field equations and allows TG, in particular $f(T)$ gravity to be compared with  metric  $f(R)$ gravity. This is an important aspect for many theories beyond GR \cite{Faraoni:2010pgm}. In TG,  second and fourth order  contributions can be easily decoupled
from each other unlike  metric gravity formulated by  the Levi-Civita connection; in this framework, $f(T,B)$ models are useful to study further degrees of freedom beyond GR (or TEGR). On the other hand, $f(T,B)$ gravity has shown promising giving  viable models at various scales, ranging from Solar System  and weak field regime \cite{Bahamonde:2016grb,Farrugia:2020fcu,Capozziello:2019msc} up to  cosmological scales \cite{Capozziello:2018qcp, Farrugia:2018gyz,Bahamonde:2020lsm,Bahamonde:2020bbc,Kadam:2022lxt,Briffa:2022fnv}. 

In TG (and, specifically in TEGR and   in its extensions), dynamical variables are not the metric components $g_{\mu \nu}$, but tetrads, $h_{a}(x^{\mu})$. Called also {\it vierbeins}, they are orthonormal vector fields defining a basis at any point $p$ of the space-time manifold $\mathcal{M}$. One can express the tetrad basis $\{ h_{a}\}$ and its dual $\{ h^{a}\}$ in terms of the holonomic coordinate basis $\{ e_{\mu} \} = \{ \partial_{\mu} \}$ and its dual  $\{ e^{\mu} \} = \{ dx^{\mu} \}$, yielding $h_{a}=h\indices{_a ^\mu}e_{\mu}h^{a}=h\indices{^a _\mu}e^{\mu} $. In this way, tetrads allow going from the space-time  manifold to the Minkowski space through 
\begin{equation}
g_{\mu \nu}=\eta_{a b}h\indices{^a _\mu}h\indices{^b _\nu}, \; \; \; \eta_{a b}=g_{\mu \nu}h\indices{_a ^\mu}h\indices{_b ^\nu},
\end{equation}
with the orthogonality conditions
\begin{equation}\label{eq2}
h\indices{ ^{a} _{\mu}} h\indices{_{b}^{\mu}}=\delta_{a}^{b}, \quad h\indices{^{a}_{\mu}} h\indices{_{a}^{\nu}}=\delta_{\mu}^{\nu}.
\end{equation}
The teleparallel connection can then be defined as  \cite{Capozziello:2018qcp}
\begin{equation}\label{eq3}
    \Gamma\indices{^{\sigma}_\nu _\mu}:=h\indices{_{a} ^{\sigma}} \partial_{\mu} h\indices{^{a} _{\nu}}+h\indices{_{a} ^{\sigma}} \omega\indices{^a _b _\mu} h\indices{^{b} _{\nu}}
\end{equation}
where $\omega \indices{^a _b _\mu}$ is the spin connection. With this connection, one can show $\nabla_{\mu} h\indices{^a _\nu} = 0$, hence the  'teleparallelism'. A particular realization of Eq.~(\ref{eq3}) is the Weitzenb\"{o}ck connection, which implies a  zero curvature, $R=0$, and a vanishing Lorentz connection, i.e. $\omega \indices{^a _b _\mu}=0$ (see \cite{Capozziello:2022zzh} for details). It is defined as 
\begin{equation}
    \Dot{\Gamma}\indices{^\rho _\mu _\nu }:= h\indices{_a ^\rho}\partial_{\nu}h\indices{^a _\mu} = - h\indices{^a _\mu} \partial_{\nu}h\indices{_a ^\rho},
\end{equation}
which is compatible both with metricity and teleparallelism conditions, $\Dot{\nabla}_{\rho}g_{\mu\nu}=0=\Dot{\nabla}_{\rho}h\indices{^a _\mu}$, where the overdot  means the derivative  w.r.t. the  Weitzenb\"{o}ck connection.  The torsion tensor is defined as 
\begin{equation}
T\indices{^\rho _\mu _\nu}:=\Dot{\Gamma}\indices{^\rho _\nu _\mu } - \Dot{\Gamma}\indices{^\rho _\mu _\nu }\,,
\end{equation}
and it is clearly antisymmetric in the last two indices. The difference between the Levi-Civita connection (designated with a ring) and the Weitzenb\"{o}ck one is given by the {\it contortion},
\begin{equation}
K\indices{^\rho _\mu _\nu} := \Dot{\Gamma}\indices{^\rho _\mu _\nu} - \mathring{\Gamma}\indices{^\rho _\mu _\nu} = -\dfrac{1}{2} \Big( T\indices{^\rho _\mu _\nu} -T\indices{_\nu ^\rho _\mu} + T\indices{_\mu _\nu ^\rho} \Big),
\end{equation}
which is instead antisymmetric on the first two indices. Finally, the torsion scalar is defined as 
\begin{equation}\label{eq6}
T := \dfrac{1}{4} T\indices{^\rho ^\mu ^\nu} T\indices{_\rho _\mu _\nu} +  \dfrac{1}{2} T\indices{^\rho ^\mu ^\nu} T\indices{_\nu _\mu _\rho} - T\indices{^\rho _\mu _\rho} T\indices{^\nu ^\mu _\nu},   
\end{equation}
which  can be shortened to $T=S\indices{^\rho ^\mu ^\nu} T\indices{_\rho _\mu _\nu}$, where
\begin{equation}
S\indices{^\rho ^\mu ^\nu}= \dfrac{1}{2}\Big( K\indices{^\mu ^\nu ^\rho} -g^{\rho \nu}T\indices{^\sigma ^\mu _\sigma} + g^{\rho \mu}T\indices{^\sigma ^\nu _\sigma}  \Big)
\end{equation}
is the so-called {\it superpotential}. Imposing the zero curvature condition, $\Dot{R}\indices{^\rho _\mu _\nu _\lambda}=0$ and contracting, one finds for the Ricci scalar in the Levi-Civita connection
\begin{equation}\label{eq8}
    \mathring{R} = -2\nabla^{\rho} S\indices{^\mu _\rho _\mu} -4\nabla^{\rho} T\indices{^\sigma _\rho _\sigma} -2 S\indices{^\rho ^\sigma ^\nu}K\indices{_\sigma  _\rho _\nu},
\end{equation}
where all covariant derivatives are Levi-Civita.  Using $S\indices{^\mu _\rho _\mu}=-T\indices{^\mu _\rho _\mu}$ and noting that the last term of Eq.~(\ref{eq8}) is equal to $-T$, one obtains the important relation
\begin{equation}
    \Dot{R} = -T -2 \nabla_{\mu}T^{\mu}
\end{equation}
where  $T^{\mu}$  indicates the contraction $T\indices{^\nu ^\mu _\nu}$. Being $h:= det (h\indices{^a _\rho})$,  we can identify the last term with the boundary term
\begin{equation}
    B= \dfrac{2}{h}\partial_{\mu}(h T \indices{^\sigma _\sigma ^\mu})
\end{equation}
thus yielding $R = -T + B$. The reader has to pay attention to the definition of $B$: in some papers, like in\cite{Capozziello:2018qcp}, the opposite sign is adopted. From now on, we omit to write rings and dots, thus implying that all quantities are calculated in the Levi-Civita connection. \\
Let us now consider the total action
\begin{equation}\label{eq11}
    \mathcal{S}_{}=\frac{1}{2 \kappa^{2}} \int d^{4} x h f(T, B)+\int d^{4} x  h  \mathcal{L}_{m}
    \end{equation}
where $\kappa^{2}=8 \pi G$, $\mathcal{L}_{m}$ is the standard matter Lagrangian, and $e=\sqrt{g}$ is the metric determinant. From the variation of the action with respect to the tetrad $h\indices{^a _\mu}$, we have   \cite{Bahamonde:2020lsm}
\begin{equation} \label{eq12}
\begin{array}{c}
-f_{T} G^{\mu}_{\nu}+\delta^{\mu}_{\nu}\Box {f_{B}}-\stackrel{}{\nabla}^{\mu} {\nabla}_{\nu} f_{B}+\frac{1}{2}\left(B f_{B}+T f_{T}-f\right) \delta^{\mu}_{ \nu} \\ + 2\left[{\nabla}^{\lambda} f_{T}+{\nabla}^{\lambda} f_{B}\right] S\indices{_\nu _\lambda ^\mu}=\kappa^{2} \mathcal{T}^{\mu}_{ \nu}
\end{array}
\end{equation}
where  $\mathcal{T}\indices{_a ^\mu} = - \frac{1}{h}\frac{\delta (h\mathcal{L}_{m} )}{\delta h\indices{^a _\mu}}$ is the matter stress-energy tensor, and we used the Einstein tensor written in the form
\begin{equation}\label{eq13}
    G_{\mu \nu} =  S\indices{^\rho ^\sigma _\mu}K\indices{_\rho _\sigma _\nu} -2 \nabla^{\rho}S\indices{_\nu _\rho _\mu} + \dfrac{1}{2}g_{\mu \nu}T
\end{equation}
in order to have a covariant form of the field equations.  In Eq.~(\ref{eq13}), we used Eq.~(\ref{eq8}) and the contracted relation  $T\indices{^\sigma ^\lambda _\sigma}=-S\indices{^\sigma ^\lambda _\sigma}$ . Notice that when $B=0$, Eq.~(\ref{eq12}) reduces to the field equations of $f(T)$ gravity, while putting $f(T,B)=f(-T+B)$ one obtains  $f(\mathring{R})$ gravity.

\section{Cosmology from $f(T,B)$ gravity}\label{sec3}
Let us now take into account cosmology from $f(T,B)$ gravity with the aim to develop a background for cosmological magnetic fields. We consider a 
(spatially) flat FRW conformal metric
\begin{equation}\label{eq14}
    ds^{2} = a^{2}(\eta)\Big( d\eta^{2} - d\mathbf{x}^2  \Big) = dt^{2} - a(t) d\mathbf{x}^2
\end{equation}
where $\eta = \int_{0}^{t} a^{-1}(t) dt $ is the conformal time and $a(t)$ is the cosmological conformal factor. A tetrad choice for this metric can be  
\begin{equation}\label{eq15}
    h\indices{^b _\mu} = a \cdot diag (1,1,1,1), \; \; \; h\indices{_b ^\mu} = \dfrac{1}{a} \cdot diag (1,1,1,1)
\end{equation}
with orthogonality  conditions given in Eq.~(\ref{eq2}). In this metric, the Ricci tensor components are
\begin{equation}
    R\indices{^i _j}= -\dfrac{1}{a^{2}}\Big( \dfrac{a''}{a} + \dfrac{{a'}^{2}}{a^{2}} \Big), \; \; \;  R\indices{^0 _0}= -\dfrac{3}{a^{2}}\Big( \dfrac{a''}{a} - \dfrac{{a'}^{2}}{a^{2}} \Big)
\end{equation}
where $i,j=1,2,3$ and prime denotes derivative w.r.t. the conformal time $\eta$. Hence, the Ricci and torsion scalars are 
\begin{equation}\label{eq17}
    R=-\dfrac{6}{a^{3}}a'', \; \; \; T=-6 \mathcal{H}^{2}
\end{equation}
where $\mathcal{H}= H/a = a'/a^{2}$ is the Hubble constant in the conformal time (see Appendix for a full computation of $T$). The boundary term then is 
\begin{equation}\label{eq18}
    B = - \dfrac{6}{a^{3}}\Big( a'' + \mathcal{H}a' a \Big).
\end{equation}
In the cosmological setup of metric~(\ref{eq14}), the field equations are
\begin{equation}\label{eq19}
\begin{array}{c}
    -f_{T} G^{0}_{0} + \dfrac{1}{a^3}\big[ 2a'f_{B}' + af_{B}'' \big] - \dfrac{1}{a^{2}} \big[ f_{B}'' - \dfrac{a'}{a}f_{B}' \big] \\
    + \dfrac{1}{2}\Big(B f_{B} + T f_{T} -f \Big) + 2 \Big( \nabla^{0} f_{T} + \nabla^{0} f_{B} \Big) S\indices{_0 _0 ^0} = \kappa^{2} \rho 
    \end{array}
\end{equation}
for the time-time component, and 
\begin{equation}\label{eq20}
\begin{array}{c}
    -f_{T} G^{i}_{j} + \delta^{i}_{j}\dfrac{1}{a^3}\big[ 2a'f_{B}' + af_{B}'' \big] + \dfrac{1}{2} \delta^{i}_{j} \big(B f_{B} + Tf_{T} - f \big) \\
 + 2 \Big( \nabla^{\lambda} f_{T} + \nabla^{\lambda} f_{B} \Big) S\indices{_j _\lambda ^i} = -\kappa^{2} p         \delta^{i}_{j}
    \end{array}
\end{equation}
for the spatial components. Here, we defined the energy density and the pressure as $\rho=T^{0}_{0}$ and $p \delta^{i}_{j} =-T^{i}_{j}$, respectively; and we used 
\begin{equation}
    \Box{f_{B}}=\dfrac{1}{a^{4}}\partial_{0}\big(a^{2}f_{B}'  \big), \; \; \; \; \nabla^{i}\nabla_{j}f_{B}= 0
\end{equation}
since $f_{B}$ may  be  a function of the conformal time only. \\
The components of the contorsion tensor in the metric ~(\ref{eq14}), from Eqs.~(\ref{eq19}) and (\ref{eq20}), are
\begin{equation}
    S\indices{_0 _\lambda ^0}=\dfrac{3}{2}\dfrac{a'}{a^3}(a^2-1)\delta^{0}_{j}, \; \; \; \; S\indices{_j _0 ^i} = -\dfrac{3}{2}\dfrac{a'}{a^3}\delta^{i}_{j}.
\end{equation}
Besides Eqs.~(\ref{eq19}) and (\ref{eq20}), the Bianchi identities for  matter,  $\nabla_{\mu}\mathcal{T}^{\mu \nu}=0$, have to be taken into account. This gives the conservation condition:
\begin{equation}\label{eq22}
 \rho' + 3 \dfrac{a'}{a}(\rho + p) = 0
\end{equation}
whose solution is  
\begin{equation}\label{eq23}
    \rho(\eta)= \Big[\dfrac{a(\eta)}{a_{0}}\Big]^{-3(1-w)} \cdot  \rho_{0}
\end{equation}
where $a_{0}$ and $\rho_{0}$ are the the scale factor and the energy density of a reference time, respectively. Notice that the above equation has the same form both in the cosmological time $t$ and in the conformal one $\eta$. \\
In the following, we assume that, during the epochs relevant for the amplification of the primordial magnetic field, i.e. de Sitter and reheating eras, the Universe is described by the action ~(\ref{eq11}) with $f(T,B)$ to be specified. We shall consider two different models in order to highlight the role of the boundary term, namely  $f(T,B)=-T + \lambda B^{n}$, which gives the Hilbert- Einstein action for $n=\{0,1\}$ (i.e. TEGR), and the non-separable model $f(T,B)=-\lambda T B^{n}$, which gives TEGR for $n=0$. In the first case the dimensions of $\lambda$ are $[\lambda]=M_{pl}^{2(1-n)}$, while in the second case $[\lambda]=M_{pl}^{-2n}$.  Notice that both these models are proven to have also  bouncing solutions, which is an alternative to the standard inflationary paradigm \cite{Caruana_2020}. Finally, they could govern the cosmological evolution today too, albeit with different values of the coupling constant $\lambda$,  in order to fit the $\Lambda$CDM phenomenology.

\subsection{The inflationary phase}
We assume that, during inflation, the energy density $\rho$ and the pressure $p$ are related by $p=-w\rho$, where $w$ is the adiabatic index. Assuming a quasi-de Sitter evolution, the scale factor is chosen as $a(\eta)=1/(-c \eta)^{\alpha}$, where $\alpha>0$, $c=H_{dS}\simeq3 \times 10^{24}$ eV \cite{Bertolami:1999}. The minus sign compensates for the negativity of $\eta$ in this epoch. Choosing $c=-1/\eta$, one finds the solution, for $\alpha$ and $w$, of the field Eqs. (\ref{eq19}) and (\ref{eq20}). 

For the model $f(T,B)=-T + \lambda B^{n}$, in the limit $\lambda=0$, the solution is  $w=1  \Longleftrightarrow \alpha = 1$, i.e. the GR (TEGR) solution is the only possible one compatible with a cosmological constant scenario.  If $\lambda/M_{pl}^{2(1-n)} \gg 1$, the general solution for $w$ is 
\begin{equation}
    w=  \dfrac{n\big[2-\alpha+8\alpha^{2} -4n(\alpha-1)^{2}\big]}{3\alpha \big[ 1-2n+2\alpha (n+1) \big]}.
\end{equation}
Having two unknowns and a single independent equation, the only way to get a solution for $\alpha$ is to choose a value of $w$. Putting $w=1$ in the above equation, one gets the double solution
\begin{equation}
 \alpha_{\pm}= \dfrac{3-5n-8n^{2}\pm 3 \sqrt{\Sigma_{1}}}{2\big(-4n^{2}+2n-6 \big)}   
\end{equation}
where $\Sigma_{1} := 16n^3-15n^2+2n+1$. Notice that, in order to avoid trivial solutions, we assume $n \not=1$ for this model, in addition to $n>0$. In this range, it results $\Sigma_1>0$. The solution  $\alpha_{-}$ is the only one compatible with GR, so we discard $\alpha_{+}$, which predicts a  scale factor with $\alpha < 1$ for all values of $n$. In particular, we have $\alpha_{-}\simeq 1$ for $n \simeq 1$ (remember that $n\not=1)$, close to the GR result, and that $\alpha_{-}<2$, $\forall n>0$. Notice that in this limit, the  torsion scalar $T$ disappears,  leaving only a boundary term power-law model. When $n>1$, then $\alpha_{-}>1$, implying a faster inflation w.r.t. GR solution. An opposite result is obtained for  $n<1$. Finally, for generic $\lambda$, it is quite difficult to obtain an analytical solution (which will clearly depend on $\lambda$) and a choice for $n$ is required.  For $n=2$, the implicit solution for $\alpha$ and a generic  $w$ is 
\begin{equation}
    \lambda = \dfrac{-2+\alpha(3w-1)}{18c^{2}(1+2\alpha)\big[-4+\alpha\big(6+4\alpha+7w-10w\alpha+\Delta\big)\big]}
\end{equation}
where $\Delta:=4(\alpha-1)(w-1)$. Finally, from Eq.~(\ref{eq23}) with $w=1$ and choosing the inflation epoch as the reference time, one finds the solution to  Eq.~(\ref{eq22}), getting $\rho(\eta)=\rho_{dS}$, where $\rho_{dS}$ is a constant depending on the specific values of $\lambda$ and $n$ (see next section for more details).  \\
For the non-minimal model $f(T,B)=-\lambda T B^{n}$, a distinction between the various regimes of the coupling constant $\lambda$ is not necessary, but the solution for a generic $w$ is very involved. For brevity, we report here the interesting case $w=1$. In this case, defining  $n_{0}=1.1617$, we have that for $ 0 \leq n \leq n_{0}$ the only positive solution is $\alpha=1$ (i.e. the GR result), while for  $  n > n_{0}$, in addition to  $\alpha_{1}=1$, one has a second positive solution, i.e.
\begin{equation}
    \alpha_{2}= \dfrac{2n^{3}+n^{2}-3n-1}{2(n^{3}+2n^{2}+3n+1)}
\end{equation}
where it is clear that $\alpha_{2}(n_{0})=0$ and $\lim\limits_{n \to \infty} \alpha_{2}(n) = 1$. 
Therefore, for this model, the range $\alpha\leq 1$, $\forall n>0$, is compatible with GR. 

\subsection{The reheating phase}
\begin{figure*}[t!!!]
\centering
\vspace{0.cm}
\includegraphics[width=1\columnwidth]{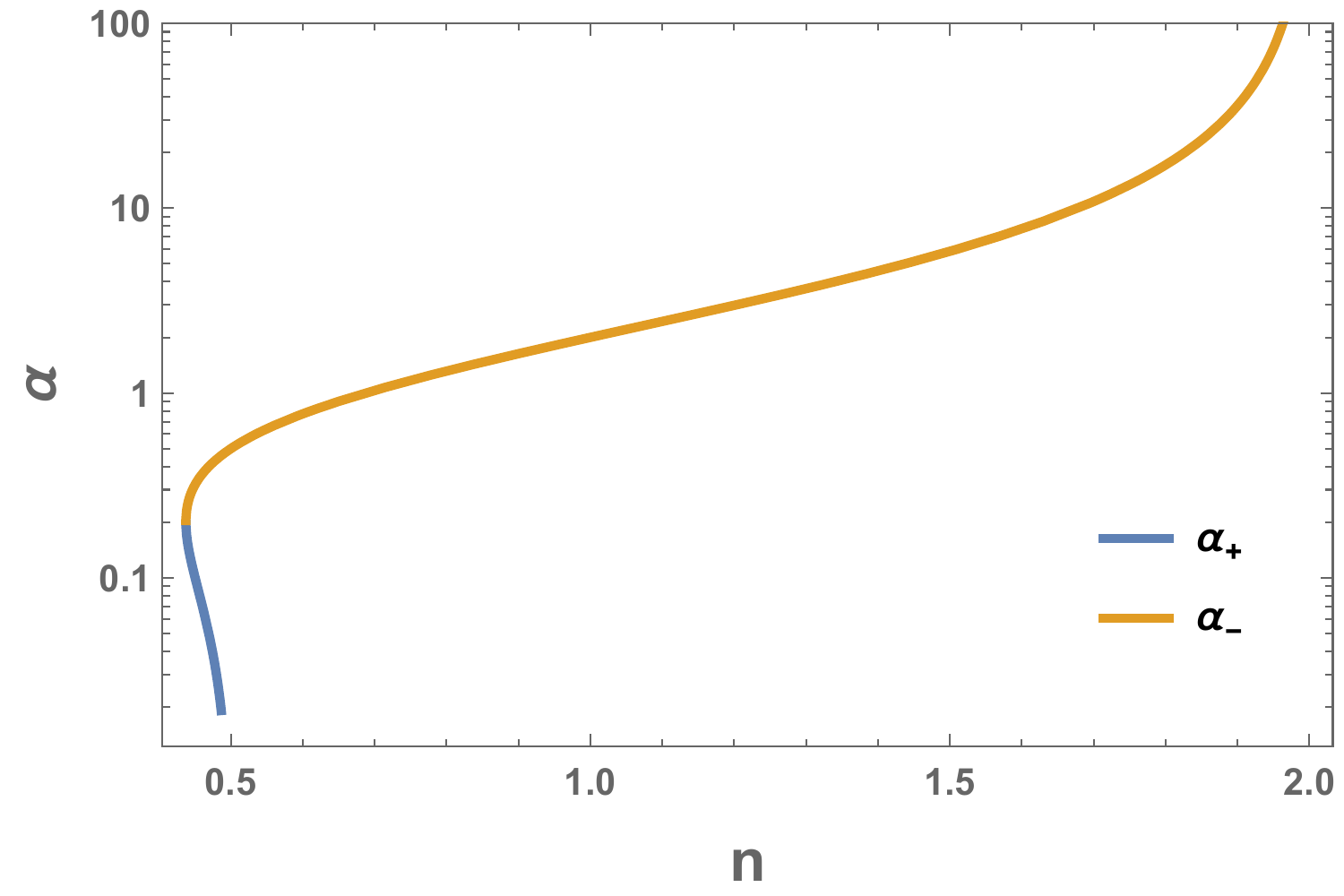}
\includegraphics[width=1\columnwidth]{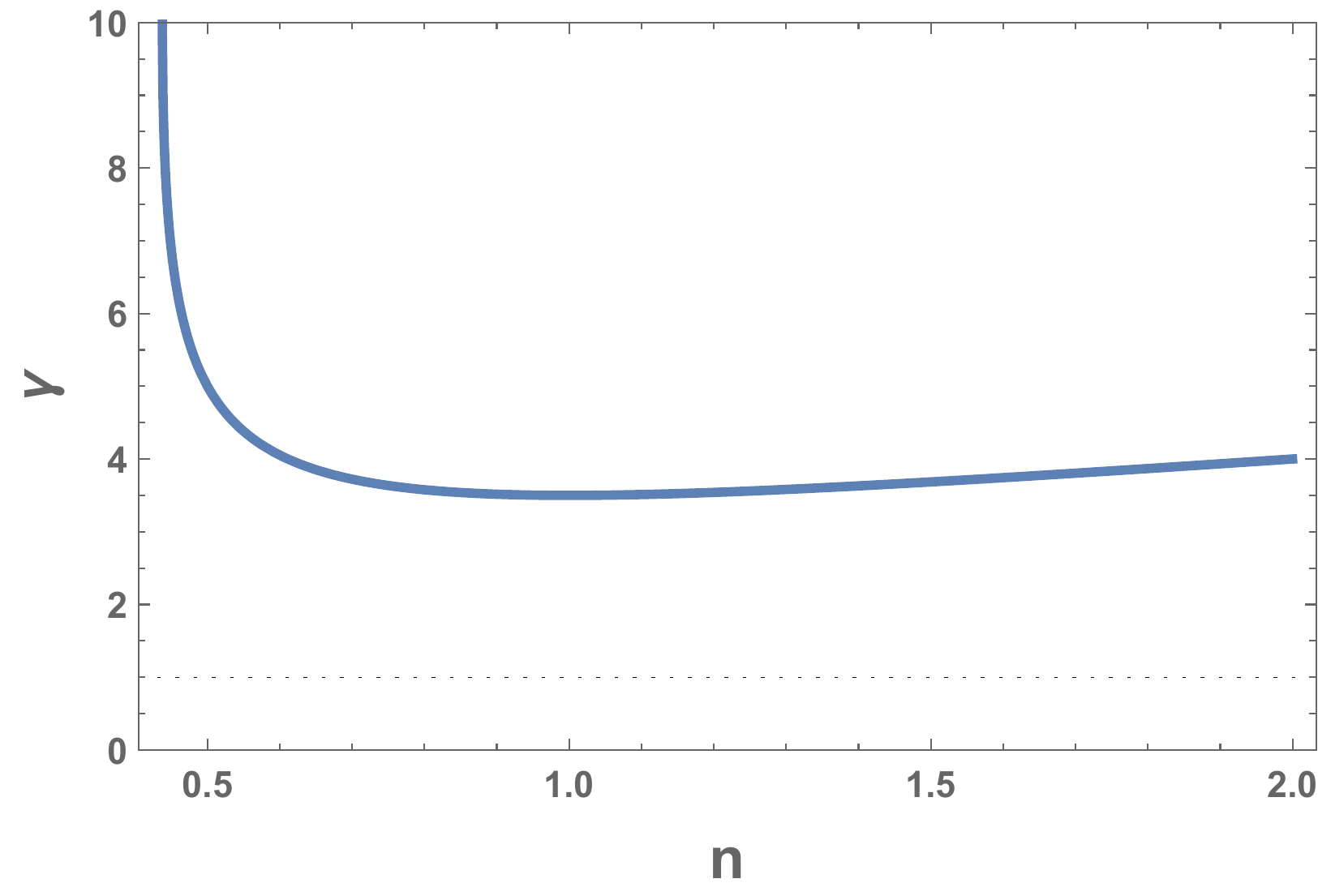}
\caption{(a) Solution (\ref{eq29}) expressing the exponent of scale factor $a$, during the reheating phase of the Universe, as a function of the power $n$ ($n\not=1$) for the model $f(T,B)=-T+\lambda B^{n}$ in the limit $\lambda/M_{pl}^{2(1-n)} \gg 1$. Notice that a solution exists only in the range $7/16 \leq n<2$. In particular, for $7/16<n<1/2$ a double solution exists (when $n=7/16$, they coincide and involve $\alpha=1/5$). Only the branch $\alpha_{-}$ is compatible with GR solution ($\alpha=2$). (b) Trend of the exponent of scalar modes $F_{k}=a^{\gamma}$ as a function of the power $n$ ($n\not=1$) for the model $f_{2}(T,B)=\lambda B^{n}$ during the reheating epoch. Notice that its value is always $>1$ (dotted line), which is the limiting case to obtain amplification w.r.t. inflationary decay. No adiabatic decay $\gamma=2$ is allowed, as we expect for a magnetic field amplification. }
\label{fig:1}
\end{figure*}

During this era, the scale factor is $a(\eta)=c^{\alpha}\eta^{\alpha}$, with  $\eta>0$ and $c=(1/4)M_{pl}^{2}H_{0}^{2}R_{0}^{3}$, where $R_{0}\sim 10^{26}h_{0}^{-1}$ m ($h_{0}\simeq 0.7$) is the present Hubble radius of the Universe and $H_{0}\sim 100 h_{0}$ km$\cdot$Mpc$^{-1}$ is the Hubble parameter. Since the pressure is zero, we take $w=0$ in  Eq.~(\ref{eq23}). Choosing $\eta=1/c$, the field equations (\ref{eq19}) and (\ref{eq20}) can be solved in a similar way to the previous case, but with one less unknown. \\
For the model $f(T,B)=-T + \lambda B^{n}$, in particular, the Eq.~(\ref{eq20}) yield
\begin{equation}
\begin{array}{c}
\alpha c^{2}(\alpha-2)+\dfrac{\lambda(n^2-n)K^{n}}{6\alpha(2\alpha-1)}\Big[8\alpha^{2}+\alpha+2-4n(1+\alpha)^{2}\Big]=0
\end{array}
\end{equation}
where we defined $K:=\big[(-1)^{}6^{}\alpha^{}c^{2}(2\alpha-1)^{}\big]$.  As before, if $\lambda=0$ (or $n=1$), the only solution is the GR one, i.e. $\alpha=2$. On the other hand, in the limit $\lambda/M_{pl}^{2(1-n)} \gg 1$, the first term is negligible, and the (double) solution is ($n\not=\{1,2\}$):
\begin{equation}\label{eq29}
\alpha_{\pm}= \dfrac{1}{8}\Big(\dfrac{1-8n\pm 3\sqrt{16n-7}}{n-2}\Big)
\end{equation}
provided that $n\geq 7/16$, in addition to the solution $\{\alpha=1/2, \forall n>0\}$.  By requiring $\alpha_{\pm}>0$, we definitely found that a solution exists only in the range $7/16 \leq n<2$, as it is shown in Fig. \ref{fig:1} (a). In particular, for $7/16<n<1/2$, a double solution exists (when $n=7/16$ they coincide and involve $\alpha=1/5$). However, as before, we will consider just the branch compatible with GR, i.e. $\alpha_{-}$. \\
For intermediate $\lambda$ values, an analytical expression for the expansion exponent $\alpha(n)$ is available only when $n$ is fixed. For  $n=2$, the implicit form for $\alpha$ is 
\begin{equation}
    \lambda = \dfrac{\alpha-2}{36c^{2}(2\alpha-1)(5\alpha+2)}.
\end{equation}
Finally, putting $w=0$ into Eq.~(\ref{eq23}) and choosing the reheating epoch as the reference time, one finds $\rho(\eta)=\rho_{RH}/a^{3}$, where $\rho_{RH}$ can be obtained from Eq.~(\ref{eq19}). In details, in the boundary term dominated regime, one gets
\begin{equation}
    \rho_{RH}= \dfrac{\lambda}{\kappa^{2}}  \big[6\alpha c^{2}(1-2\alpha)\big]^{n}\Big[ \dfrac{1}{2}(n-1)+(\alpha+1)\dfrac{n(n-1)}{(2\alpha-1)} \Big]
\end{equation}
where $\alpha$ is here expressed by $\alpha_{-}$, so to get an expression which depends only on the  model parameter $n$, as well as on the constants $\lambda$, $\kappa$ and $c$. 

For the non-minimal model $f(T,B)=-\lambda T B^{n}$, finding  a solution for $\alpha$ from Eq.~(\ref{eq20}) is more challenging. As before, there is the constant solution $\alpha=1/2$, $\forall n>1$. The equation is now of  fourth degree; however, only two solutions are real and positive. We define them $\alpha_{1,2}$ and show some values in the Table~\ref{tab1}. When $n=0$, the only solution is $\alpha=2$, i.e. the GR limit. As it can be seen from the table, $\alpha_{1}<1/2$ for every $n$; therefore the branch to be taken is $\alpha_{2}$, which is also in agreement with the GR solution. Furthermore, for $n>3$, just one (positive) solution exists, namely $\alpha_{2}$. Finally, from Eqs.~(\ref{eq23}) and (\ref{eq19}), one finds the expression for the energy density, i.e.
\begin{equation}
    \rho_{RH}= \dfrac{\lambda(-1)^{n}6^{n}c^{2n+2}}{\kappa^{2}}\Omega_{RH}
\end{equation}
where we defined
\begin{equation}
    \Omega_{RH}:= \dfrac{3 \alpha^{n+2}}{(2\alpha-1)^{1-n}}\Big[ \alpha(2\alpha-1) +2n\Big(\alpha^{}n+ n +\alpha- \dfrac{1}{2} \Big) \Big]
\end{equation}
where the values of $\alpha$ are given by $\alpha_{2}$ in Table~\ref{tab1}, so to obtain a relation depending only on the power $n$. 

\begin{table}[b]
\caption{\label{tab1}Solutions of  Eq.~(\ref{eq20}) for the expansion exponent $\alpha$ in the reheating era, for the non-minimal model $f(T,B)=-\lambda T B^{n}$ as a function of the power $n$. Although the equation is of  fourth degree, only two are real and positive. The branch compatible with GR is only $\alpha_{2}$. In the last column, the exponent of the Fourier mode $F_{k}\sim a^{\gamma}$ is reported, using the $\alpha_{2}$ solution.  }
\begin{ruledtabular}
\begin{tabular}{lccc}
\textrm{n}&
\textrm{$\alpha_{1}$}&
\textrm{$\alpha_{2}$}&
\textrm{$\gamma$}
\\
\colrule
1/2 & 0.35 & 1.81 & 5.21\\
1 & 0.17 & 1.70 & 6.90\\
3/2 & 0.03 & 1.74 & 8.45\\
2 & 0.08 & 1.81 & 9.87\\
5/2 & 0.18 & 1.89 & 11.23\\
3 & -0.25 & 1.97 & 12.57\\
7/2 & -0.32 & 2.05 & 13.88\\
\end{tabular}
\end{ruledtabular}
\end{table}

\section{The non-minimal coupling}\label{sec4}
Let us now add a non-minimal coupling to the action  between matter (photons) and torsion, in order to break the conformal invariance\footnote{It is possible to overcome electromagnetic conformal invariance in a \textit{minimal} contest too. The main idea in this case is to generalize the electromagnetic Lagrangian and promote it to a non-linear function of $F\doteq (1/4)F_{\mu \nu}F^{\mu nu}$. See for example \cite{Bertolami_2022,MosqueraCuesta:2009tf,MosqueraCuesta:2017iln,Otalora_2018,Dittrich_1998} and references therein.} which leads to a fast decay of any magnetic field generated in the  primordial Universe. There are various ways to do this, both tensorial and scalar couplings, as for $f(R)$ theories. GR assumes that the electromagnetic stress-tensor curves the space-time as any other source of energy density. However, there are strong reasons to expect that this picture should be changed for very strong gravitational fields, where non-minimal couplings  should appear.  On the one hand, it is possible that, at the very high energies of early Universe, all forces were unified and described by a single field then decayed through phase transitions (the inflation itself could be a product of a phase transition of this field \cite{Watari_2004}). On the other hand, QED calculations \cite{Drummond:1979pp} on the vacuum polarization in curved space-times show a  non-minimal coupling between gravity and electromagnetism, coming from the  tidal influences of the space-time geometry on the production of electron/positron pairs in the vacuum.\\
The action ~(\ref{eq11}) can be then modified as
\begin{equation}\label{eq32}
    \int    \Big[ \dfrac{1}{2\kappa^{2}} f_{1}(T,B) +  ( f_{2}(T,B)+1 ) \mathcal{L}_{m}  \Big]\, e \, d^{4}x
\end{equation}
where $f_{1}$ and $f_{2}$ are two sufficiently smooth arbitrary functions of the scalar torsion $T$ and the boundary term $B$, and $e=a^{4}(\eta)$. As we are interested in a coupling between gravity and magnetic fields, we take $\mathcal{L}_{m}= -\frac{1}{4}F^{\mu\nu}F_{\mu\nu}$, where the electromagnetic tensor $F_{\mu\nu}$, in the metric Eq.~(\ref{eq14}), is given by 
\begin{equation}\label{eq33}
F_{\mu \nu}=a^{2}(\eta)\left(\begin{array}{cccc}
0 & E_{x} & E_{y} & E_{z} \\
-E_{x} & 0 &  -B_{z} & B_{y} \\
-E_{y} & B_{z} & 0 &  -B_{x} \\
-E_{z} & -B_{y} & B_{x} & 0 
\end{array}\right)
\end{equation}
thanks to conformal invariance of classical electromagnetism. The full contravariant components $F^{\mu\nu}$ are obtained from those of Eq.~(\ref{eq33}), changing the sign to the electric field $\mathbf{E}$ and replacing $a^{2}$ with $1/a^{2}$. Notice that the definition of $F$  depends on the metric signature: if the signature is $(-,+,+,+)$, \textit{all} signs should be changed. Calling $\Tilde{F}_{\mu\nu}$ the piece in brackets, corresponding to the  electromagnetic fields  as measured by a comoving (inertial) observer, and $F_{\mu\nu}=\partial_{\mu}A_{\nu}-\partial_{\nu}A_{\mu}$ the Faraday tensor in a curved space-time, then the following relations for every couple of spatial indices $\{i,j\}$ fixed hold:
\begin{equation}
    F^{ij}=\dfrac{1}{a^{4}}F_{ij}=\dfrac{1}{a^{4}} \big( \partial_{i}{A}_{j}-\partial_{i}{A}_{j} \big)=\dfrac{1}{a^{2}}\Tilde{F}_{ij}\,,
\end{equation}
where ${A}_{\mu}= ({\Phi},-{\mathbf{A}})$ is the gauge $U(1)$ field in the curved space-time (a minus sign in the above relations appears for the other non-zero components $F^{j 0}$, regardless of the metric signature). To clarify, we highlight that $\Tilde{F}_{\mu\nu}$ is not the Minkowski Faraday tensor: generally, the electric and magnetic fields measured by
inertial (here comoving) observers in a conformally flat space-time  does not coincide with their values
in Minkowski space-time as these frame are not equivalent.  Here, a comment is in order. In Eq.~(\ref{eq32}), the above $\lambda$ can play the role of coupling constant  with photons. Indeed, if these couplings played a dominant role in the primordial Universe, at the present epoch, they could be very small for cosmological background. In the case of a Riemann tensor coupling of the form $\lambda R^{\mu\nu\rho\sigma}F_{\mu\nu}F_{\rho\sigma}$, for example, it has been shown \cite{Prasanna:2003ix} that $\lambda < 10^{22}$ eV$^{-2}$ today. In this section, we take the models $f_{2}(T,B)=\lambda B^{n}$ and $f_{2}(T,B)=-\lambda TB^{n}$, with the only constraint $n\not= 0$. As in the previous section, the dimensions of $\lambda$ will depend on the specific power $n$. \\
The energy-momentum tensor corresponding to the action (\ref{eq32}) is 
\begin{equation}
    T^{\mu\nu} = (1+f_{2})\big[ - F^{\mu\sigma}F\indices{^\nu _\sigma}+\dfrac{1}{4}F^{\alpha\beta}F_{\alpha\beta}g^{\mu\nu} \big],
\end{equation}
where, again,  we are using natural units. \\
Varying the action (\ref{eq32}) w.r.t. the conformal potential vector $A_{\mu}$, one gets the field equations
\begin{equation}\label{eq35}
\partial_{\mu}\Big[ e \Big( F^{\mu\nu} + f_{2}(T,B)F^{\mu\nu} \Big)\Big] = 0
\end{equation}
where we decided to maintain the first term coming from classical electromagnetism, differently from \cite{Bertolami:2022hjk}. Classical electromagnetism is obtained when $f_{2}=0$ is taken into account. In the FRW background (\ref{eq14}), the electromagnetic potential vector $A_{j}$ satisfies, in the conformal time, the following differential equations: 
\begin{equation}\label{eq36}
    \Big[ A_{j}''(\eta,\mathbf{x}) +  \dfrac{f_{2}'}{1+f_{2}} A_{j}'(\eta,\mathbf{x}) -\Delta A_{j}(\eta,\mathbf{x}) \Big]\delta^{j\nu} = 0
\end{equation}
when $\nu=j$, where $j=1,2,3$ and we defined $\Delta:=\delta^{ki}\partial_{k}\partial_{i}$ (where $k,i=1,2,3$); and 
\begin{equation}
\Big[\delta^{ij} \Big(1+f(T,B)\Big) \partial_{i}\big( \partial_{0}A_{j} \big) \Big] \delta^{\nu}_{0} = 0
\end{equation}
valid when $\nu=0$. Here, we used the radiation gauge, i.e. $A_{0}(t,\mathbf{x})=0$ and $\partial_{j}A^{j}(t,\mathbf{x})=0$ (compatible with zero net charge). In particular, the above equation is nothing more than the null diverge condition $\nabla \cdot \mathbf{E} = 0$, as we expect. This condition  can also be obtained directly by deriving the gauge fixing condition. Notice that  this does not imply a null electric field (see later).  Using  the relations 
\begin{equation}
A_{i}'=a^{2}E_{i}\;, \; \; \; \; B_{i}=-a^{-2}\epsilon_{ijk}\partial_{j}A_{k} 
\end{equation}
where $\epsilon$ is the totally anti-symmetric Levi-Civita symbol (implicit summation over repeated indices), then Eq.~(\ref{eq36}) can be written in function of the electric and magnetic fields as
\begin{equation}\label{eq37}
    \partial_{0}\Big[ a^2 E_{k} \Big(1 + f_{2}(T,B)\Big) \Big] + a^{2}\Big(1+f_{2}(T,B)\Big)\partial_{j}\big(\epsilon_{jki}B_{i} \big) = 0.
\end{equation}
Similarly, the dual tensor $F^{*\mu\nu}=\frac{1}{2}F_{\alpha\beta}\epsilon^{\alpha\beta\mu\nu}$ obeys \footnote{To obtain the complete set of electromagnetic equations, the Bianchi identities must be considered.}
\begin{equation}
    \nabla_{\mu} \Big( (-g)^{-\frac{1}{2}} F^{*\mu\nu} \Big) = 0 \; \; ,
\end{equation}
which can be recast as 
\begin{equation}\label{eq38}
 \partial_{0}\Big[ a^2 B_{k}  \Big] - a^{2}\partial_{j}\big(\epsilon_{jki}E_{i} \big) = 0.   
\end{equation}
together with the condition $\nabla \cdot \mathbf{B}=0$.  Eqs.~(\ref{eq37}) and (\ref{eq38}) are the Maxwell equations in the curved space-time (\ref{eq14}) derived from the  non-minimal coupling action (\ref{eq32}). One easily verifies that when $f_{2}(T,B)=0$, the usual Maxwell equations in GR are recovered. Deriving Eq.~(\ref{eq37}) w.r.t spatial indices, multiplying by $\epsilon^{\lambda k l}$ and using Eq.~(\ref{eq38}), one finds a purely magnetic equation, i.e.
\begin{equation}\label{eq39}
    \partial_{0}^{2}\Big[a^{2} \Big(1+ f_{2}(T,B)\Big) B_{k}   \Big] - a^{2}\Big(1+ f_{2}(T,B)\Big)\Delta B_{k} = 0
\end{equation}
where we used the relation
\begin{equation*}
 \partial_{\lambda}\partial_{j}\Big(\epsilon\indices{_\lambda _k _l} \epsilon\indices{_j _k _i} B_{i} \Big) = \delta^{ij}\partial_{i}\partial_{j} B_{l}\; \;,
 \end{equation*}
 and we assumed that 
 \begin{equation*}
     \dfrac{\partial }{\partial \tau}\Big(a^2 B_{k}f_{2}(\ln{f_{2}})' \Big) \simeq 0
 \end{equation*}
since $(\ln{f_{2}})'\equiv \partial_{0}(\ln{f_{2}})\simeq 0$. 
In this way, we have eliminated explicitly the electric field, but it is still present as derivative of $\mathbf{B}$ in the first term of Eq.~(\ref{eq39}).  
Since $\mathbf{B}(\eta,\mathbf{x})=a^{-2}\nabla \times \mathbf{A}$ and defining the  spatial Fourier transform of the magnetic field vector
\begin{equation}
\mathbf{B}(\eta,\mathbf{k})=\int \dfrac{d^{3}\mathbf{x}}{2\pi} e^{i \mathbf{k}\cdot \mathbf{x}} \mathbf{B}(\eta,\mathbf{x}),
\end{equation}
then Eq.~(\ref{eq39}) becomes 
\begin{equation}\label{eq41}
    \partial_{0}^{2}\Big[\Big(1+f_{2}(T,B)\Big) \mathbf{F}_{k} \Big] + k^{2}\Big(1+f_{2}(T,B)\Big) \mathbf{F}_{k} = 0,
\end{equation}
i.e. the second order differential equation 
\begin{equation}\label{eq42}
\begin{array}{cc}
\mathbf{F}_{k}'' \Big(1 + f_{2}(T,B)\Big) + 2 f_{2}'(T,B) \mathbf{F}_{k}' \\ + \Big[f_{2}''(T,B) + k^{2} f_{2}(T,B) + k^{2} \Big] \mathbf{F}_{k} = 0
\end{array}
\end{equation}
where we defined
\begin{equation}\label{eq43}
\mathbf{F_{k}}(\eta):=a^{2}(\eta)\mathbf{B}(\eta,\mathbf{k}) \end{equation}
with $k=|\mathbf{k}|$ (here k is not an index). When $f_{2}=0$, then  from Eqs.~(\ref{eq38}) and (\ref{eq39})  it follows that 
\begin{equation*}
    |\mathbf{B}| \propto \dfrac{1}{a^{2}(\tau)}\;\; , \; \; |\mathbf{E}| \propto \dfrac{1}{a^{2}(\tau)}
\end{equation*}
 which is the adiabatic trend\footnote{It is not correct to derive exact solutions for the fields directly from Eq.~(\ref{eq42}) since it is not fully equivalent to  Eqs.~(\ref{eq38}) and (\ref{eq39}). Indeed, to arrive at Eq.~(\ref{eq42}), we derived one more time. By solving it, one would get a physical equivalent solution but mathematically different, namely $|\mathbf{B}| \propto 1/a^{3}$. }  in the standard GR settings. As during inflation $a(\tau) \rightarrow \infty$, this would mean a strong and fast decay, preventing any form of amplification.
Notice that the module $F_{k}:= |\mathbf{F}_{k}|=\sqrt{\mathbf{F}_{k}^{*}\mathbf{F}_{k}}$ is a good approximation of the magnetic flux through the expanding Universe. \\
A comment here is in order. If the electric field is assumed to be zero, i.e. $F^{0j}=0$, then just one of the two Maxwell equations is required, namely Eq.~(\ref{eq38}). Indeed, imposing $\mathbf{E}=0$, deriving w.r.t. to the conformal time $\eta$ and Fourier transforming, this equation gives
\begin{equation}
\mathbf{F}_{k}'' \Big(1 + f_{2}(T,B)\Big) + 2 f_{2}'(T,B) \mathbf{F}_{k}' + f_{2}''(T,B)\mathbf{F}_{k} = 0
\end{equation}
which differs from Eq.~(\ref{eq42}) only for the $k$-terms. Effectively, these terms are null as evident solving the other Maxwell equation, Eq.~(\ref{eq37}), which gives the vector equation 
\begin{equation}
    a^{2} \Big(1 + f_{2}(T,B)\Big) \nabla \times \mathbf{B}(\eta,\mathbf{x}) = 0
\end{equation}
which, after a Fourier transform (and multiplication by $i \mathbf{k}$), becomes $(1+f_{2})k^{2}\mathbf{F}_{k} = 0$. Since $F_{k}=0$ would be a trivial solution, this means that imposing a zero electric field, automatically leads to the super-horizon approximation $k \eta \leq 1$ in the Maxwell equations. However, the inverse is not generally true. Indeed, imposing this approximation to the Maxwell Eqs. (\ref{eq37}) and (\ref{eq38}) (after Fourier transforming), gives $E \propto 1/a^{2}$ (as well as for $\mathbf{B}$) during inflation (where we neglect the non-minimal coupling function $f_{2}$), and $E \propto 1/(a^{2}f_{2})$ during the reheating epoch. Therefore, if during inflation one could state that $E\approx 0$ (the scale factor grows exponentially), the trend in the reheating era depends on the specific function $f_{2}$, and so $E\not= 0$ in general. However, assuming a zero electric field from the beginning is not a natural choice, since the context is that of a quickly varying magnetic flux.    \\
In order to evaluate the magnetic field for the de Sitter and reheating phases of the Universe, we concern ourselves  with the evolution of the magnetic field fluctuations whose wavelengths are well outside the horizon, i.e. we assume $k \eta \ll 1$. Only this condition can ensure large scale magnetic fields. Furthermore, we will assume a strong non-minimal coupling only in the reheating era.  Generally, in order to have amplification, if the magnetic flux decays as $a^{\gamma}$, then it should be $\gamma_{rh}> - \gamma_{dS}$, where $\gamma_{dS,rh}$ stay for the power during inflation and reheating epochs, respectively.  

\subsection{The inflationary phase}

As a first case, we consider the purely boundary term model $f_{2}(T,B)=\lambda B^{n}$. During this phase, we assume that  the non-minimal coupling term is zero (or  negligible). Therefore, Eq.~(\ref{eq42}) gives  the harmonic oscillator differential equation 
\begin{equation}\label{eq44}
\mathbf{F}_{k}''(\eta) + k^2 \mathbf{F}_{k}(\eta) = 0
\end{equation}
whose module solution  is
\begin{equation}
    F_{k}=\sin(k\eta)\sim \dfrac{1}{a}
\end{equation}
where we used the super-horizon approximation $k \eta \ll 1$. Here, we assumed $\alpha=1$ (as well as $w=1$), as found in the previous section for this regime. Therefore, no divergences from GR appear in this epoch.  \\

For the non-minimal model $f_{2}(T,B)=-\lambda T B^{n}$, making the above assumptions, one arrives at the same Eq.(\ref{eq44}), and then the same solution $F_{k}\sim \eta$. Hence,  as long as $n\leq n_{0}$, one simply have $\gamma_{dS}=-1$. Otherwise, for the alternative solution $\alpha<1$, the condition becomes $\gamma_{dS}<-1$, thus requiring a strong constraint  on the $\gamma$ power of the reheating era in order to have possible amplifications. For example, when $n=3$, then $\alpha \simeq 1/2$. In this case, to have an amplification effect, one should expect a power-law solution like $a^{\gamma}$ with $\gamma > 2$ in the reheating epoch. 

\subsection{The reheating phase}

During this phase, we assume dominant the non-minimal coupling. Therefore, Eq.~(\ref{eq42}) becomes 
\begin{equation}
\begin{array}{c}
    \mathbf{F}_{k}'' + 2nB^{-1}B'\mathbf{F}_{k}' \\
    + \Big[n(n-1)(B')^{2}B^{-2}+nB^{-1}B''+k^2+\dfrac{k^2}{\lambda B^{n}} \Big] \mathbf{F}_{k} = 0
    \end{array}
\end{equation}
where $B$ is the boundary term of Eq.~(\ref{eq18}). Writing $\mathbf{F}_{k}$ as a function of the scale factor $a$, hence using the relations 
\begin{equation}
    \begin{array}{ll}
    \mathbf{F}_{k}'(\eta) = \dfrac{d\mathbf{F}_{k}}{da} a' \; ,\\
         \mathbf{F}_{k}''(\eta) = {a'}^{2}\dfrac{d^{2}\mathbf{F}_{k}}{da^{2}} + \dfrac{d\mathbf{F}_{k}}{da}\Big(\dfrac{\alpha-1}{\alpha} \Big) \dfrac{{a'}^{2}}{a} 
         
    \end{array}
\end{equation}
one gets 
 \begin{equation}
\mathbf{F}_{k}''(a) + \mathcal{C}(n) \mathbf{F}_{k}'(a) + \dfrac{\mathcal{D}(n)}{a^{2}} \mathbf{F}_{k}(a) = 0
\end{equation}   
where the prime  denotes derivative w.r.t. the new variable $a$ and we define 
\begin{equation}
 \mathcal{C}(n):=\dfrac{\alpha-1}{\alpha}-4n\Big(\dfrac{\alpha+1}{\alpha} \Big)
 \end{equation}
 and 
 \begin{equation}
     \mathcal{D}(n):=\dfrac{4n(n-1)(\alpha+1)^{2}}{\alpha^{2}} + \dfrac{2n(3+2\alpha)(\alpha+1)}{\alpha^{2}}.
 \end{equation}
In the above functions one should substitute $\alpha_{-}$ from Eq.~(\ref{eq29}). With the ansatz $F_{k}(a)=a^{\gamma}$, the following solution is found
\begin{equation}\label{eq48}
    F_{k}(a)=c_{1} a^{\frac{1}{2}\big(-\sqrt{\mathcal{G}(n)}-\mathcal{C}(n)+1\big)}+c_{2}a^{\frac{1}{2}\big(\sqrt{\mathcal{G}(n)}-\mathcal{C}(n)+1\big)}
\end{equation}
where $c_{1,2}$ are constants and 
\begin{equation}\label{eq49}
    \mathcal{G}(n):=\mathcal{C}^{2}(n)-2\mathcal{C}(n)-4 \mathcal{D}(n)+1 .
\end{equation}
Solution ~(\ref{eq48}) is valid as long as $\mathcal{G}(n)>0$. When $\mathcal{G}(n)=0$, i.e. $\mathcal{D}=(\mathcal{C}^{2}-2\mathcal{C}+1)/4 $,  then the solution is
\begin{equation}\label{eq50}
F_{k}(a)=c_{1} a^{\frac{1-\mathcal{C}(n)}{2}}+c_{2} \big(\mathcal{C}(n)-1\big)a^{\frac{1-\mathcal{C}(n)}{2}}\log(a).
\end{equation}
For our background solution $\alpha_{-}$, the condition $\mathcal{G}(n)=0$ is reached only in the limiting case $n \rightarrow 2$, hence we neglect it in the following and consider just the solution   (\ref{eq48}). Therefore, as showed in Fig.\ref{fig:1} (b), we always have $\gamma>1$ in all the range of existence of $\alpha_{-}$. Since the exponent is greater than the inflationary one ($\gamma_{dS}=-1$), a magnetic field amplification is always possible ($\gamma_{rh}>1$). Also, notice  that any adiabatic decrease is avoided.  \\

For the non-minimal model $f_{2}(T,B)=-\lambda T B^{n}$, instead, assuming as before a dominant non-minimal gravity-photon coupling, and rewriting in the variable $a$, Eq.~(\ref{eq42}) becomes 
\begin{equation}
\mathbf{F}_{k}''(a) + \mathcal{Q}(n) \mathbf{F}_{k}'(a) + \dfrac{\mathcal{S}(n)}{a^{2}} \mathbf{F}_{k}(a) = 0
\end{equation} 
where prime here stay for derivative w.r.t. the scale factor $a$ and we define
\begin{equation}
 \mathcal{Q}(n):= - \dfrac{\Big(5+3\alpha+4n(1+\alpha) \Big)}{\alpha}
 \end{equation}
 and
 \begin{equation}
     \mathcal{S}(n):=\dfrac{2(1+\alpha)(1+n)}{\alpha^{2}}\Big(3+2n+\alpha(1+n) \Big),
 \end{equation}
where $\alpha$ is given by $\alpha_{2}$ in Table~\ref{tab1}. The solution of this differential equation is analogous to the previous case, i.e.
\begin{equation}\label{eq52}
 F_{k}(a) \sim  a^{\frac{1}{2}\big(\sqrt{\mathcal{H}(n)}-\mathcal{Q}(n)+1\big)}  
\end{equation}
where, in a completely analogous way to Eq.~(\ref{eq49}), we define
\begin{equation}
    \mathcal{H}(n):=\mathcal{Q}^{2}(n)-2\mathcal{Q}(n)-4 \mathcal{S}(n)+1 .
\end{equation}
Denoting  by $\gamma$ the exponent
of Eq.~(\ref{eq52}), we listed it in the last column of Table~\ref{tab1}. In the special case $n=0$, the (only) solution $\alpha=2$ leads to $\gamma=3.5$, which is high enough to allow amplification w.r.t. the inflationary period. Interestingly, as the power $n$ increases, this value increases in turn, similarly  to what was found in \cite{Lambiase:2008zz} for a non-minimal coupling involving the Riemann tensor and the photon. At least, we have $\gamma>3$, then the amplification is possible also for more \textit{exotic} (meaning far from GR) inflationary solutions, like $\alpha=1/2$ or even $\alpha=1/3$.   Finally, it is worth noticing that, in this discussion, the role of the coupling constant $\lambda$ is irrelevant, since it disappears thanks to the super-horizon approximation.

\section{Amplification during inflation}\label{sec5}
In the previous sections, we assumed that the main amplification for the seed of PMs was achieved during the \textit{reheating} epoch, neglecting a gravity-photon non-minimal coupling during inflation. In this section, we generalize and extend what was found in \cite{Bertolami:2022hjk}, in order to estimate amplification of the magnetic field assuming its manifestation during \textit{inflation}, i.e. turning on the non-minimal coupling in this epoch, rather than in the reheating one. Indeed, neglecting the classical electromagnetic term in Eq.~(\ref{eq35}), and assuming a negligible electric field,  the consequent Eq.~(\ref{eq37})  ensures that the magnetic field intensity goes like
\begin{equation}\label{eq54}
|\mathbf{B}(\eta, \mathbf{x})| \propto \dfrac{1}{a^{2}f_{2}(T,B)}. 
\end{equation}
Notice that to derive the above relation, one can alternatively   use a different definition of $\mathbf{F}_{k}$, instead of Eq.~(\ref{eq43}), to brake the adiabatic decay, namely \cite{Bertolami:2022hjk}
\begin{equation}
\mathbf{F_{k}}(\eta):=a^{2} f_{2}(T,B)(\eta)\mathbf{B}(\eta,\mathbf{k})
\end{equation}
whose Fourier transform gives
\begin{equation}
    \mathbf{B}(\eta,\mathbf{x})= \dfrac{1}{a^{2} f_{2}(T,B)} \int \mathbf{F}_{k}(\eta)e^{i\mathbf{k}\cdot \mathbf{x}} d\mathbf{k}.
\end{equation}
Thanks to the super-horizon approximation, the integral is bounded, thus giving the  relation (\ref{eq54}). Notice that, in the GR limit, $f_{2}=1$, one would have $\mathbf{B} \propto 1/a^{2}$, i.e. an adiabatic decay. \\
The rate  between the magnetic field at the beginning and the end of the inflation period is 
\begin{equation}
    \dfrac{\mathbf{B}_{end}}{\mathbf{B}_{in}} \simeq \Big(\dfrac{a_{in}}{a_{end}} \Big)^{2} \dfrac{f_{2}^{in}(T,B)}{f_{2}^{end}(T,B)}
\end{equation}
where $f_{2}^{in}(T,B)$ is the non-minimal function at the beginning of inflation, while $f_{2}^{end}(T,B)$ is its evaluation at the end, which we can approximate with the beginning of the reheating epoch. Since $a_{end} \sim e^{60} a_{in}$ and considering the model $f_{2}(T,B)=\lambda B^{n}$, we have
\begin{equation}
 \dfrac{\mathbf{B}_{end}}{\mathbf{B}_{in}} \simeq 10^{-53} \Big( \dfrac{H_{in}}{H_{end}}\Big)^{2n},
\end{equation}
where we used the relations
\begin{equation}
    T = - 6H^{2}, \; \; \; B \simeq - 18 H^{2} 
\end{equation}
($H=\Dot{a}/a$) valid in the well-known slow-roll approximation, i.e.  $\Dot{H} \simeq 0$, which ensure that the inflaton evolution is sufficiently damped to allow for an accelerated expansion.  \\
At the end of inflation, the Hubble constant is  \cite{Bertolami:2022hjk}
\begin{equation}
H_{end} \simeq \dfrac{\pi}{\sqrt{90}}\dfrac{T^{2}_{RH}}{M_{pl}}    
\end{equation}
where $T_{RH}$ is the reheating temperature. During inflation, instead, it is usually assumed $H_{in} \simeq 10^{-6} M_{pl} $.
Finally, since $10^{-8} M_{pl} \lesssim T_{RH}\lesssim 10^{-4} M_{pl}$ \cite{Bertolami:1999},   we have that
\begin{equation}
10^{-53+4n} \lesssim \dfrac{\mathbf{B}_{end}}{\mathbf{B}_{in}} \lesssim 10^{-53+20n}.  
\end{equation}
Let us consider now the parameter $r = \rho_{B}/ \rho_{\gamma}$ described in Sec. \ref{sec1}. From this definition, it is easy to deduce the relation with the magnetic field intensity, namely
\begin{equation}\label{eq62}
\dfrac{\mathbf{B}_{end}}{\mathbf{B}_{in}} = \sqrt{\dfrac{r_{RH}}{r_{IN}}}.
\end{equation}
Here, we considered $\rho_{\gamma}$  constant during inflation, and we called $r_{RH,IN}$ the rate value in the reheating and inflation epochs, respectively. 

Even if we know with enough confidence that the present value is $r\simeq1$ and  $r\approx10^{-34}$ or $r\approx10^{-8}$ for the pre-galactic epoch, we have no idea of its value in other periods of the Universe evolution, so some assumptions have to be made. Following \cite{Turner:1987bw}, it is possible to consider that the pre-galactic value, $r_{pg}$, is related to $r_{RH}$ by  $r_{pg} \approx  10^{-14} r_{RH}$. From Eq.~(\ref{eq62}), and considering $r_{IN} = 10^{\chi}$ with $\chi$ an arbitrary power, the following constraints are deduced:
\begin{itemize}
    \item if $r_{pg}=10^{-34}\;\;$  $\longrightarrow$  $\; \; \chi \geq 58 - 40 n$
     \item if $r_{pg}=10^{-8}\;\;\;\,$ $  \longrightarrow$  $\; \; \chi \geq 84 - 40 n$.
\end{itemize}

\begin{figure}[t]
\includegraphics[width=8.1cm]{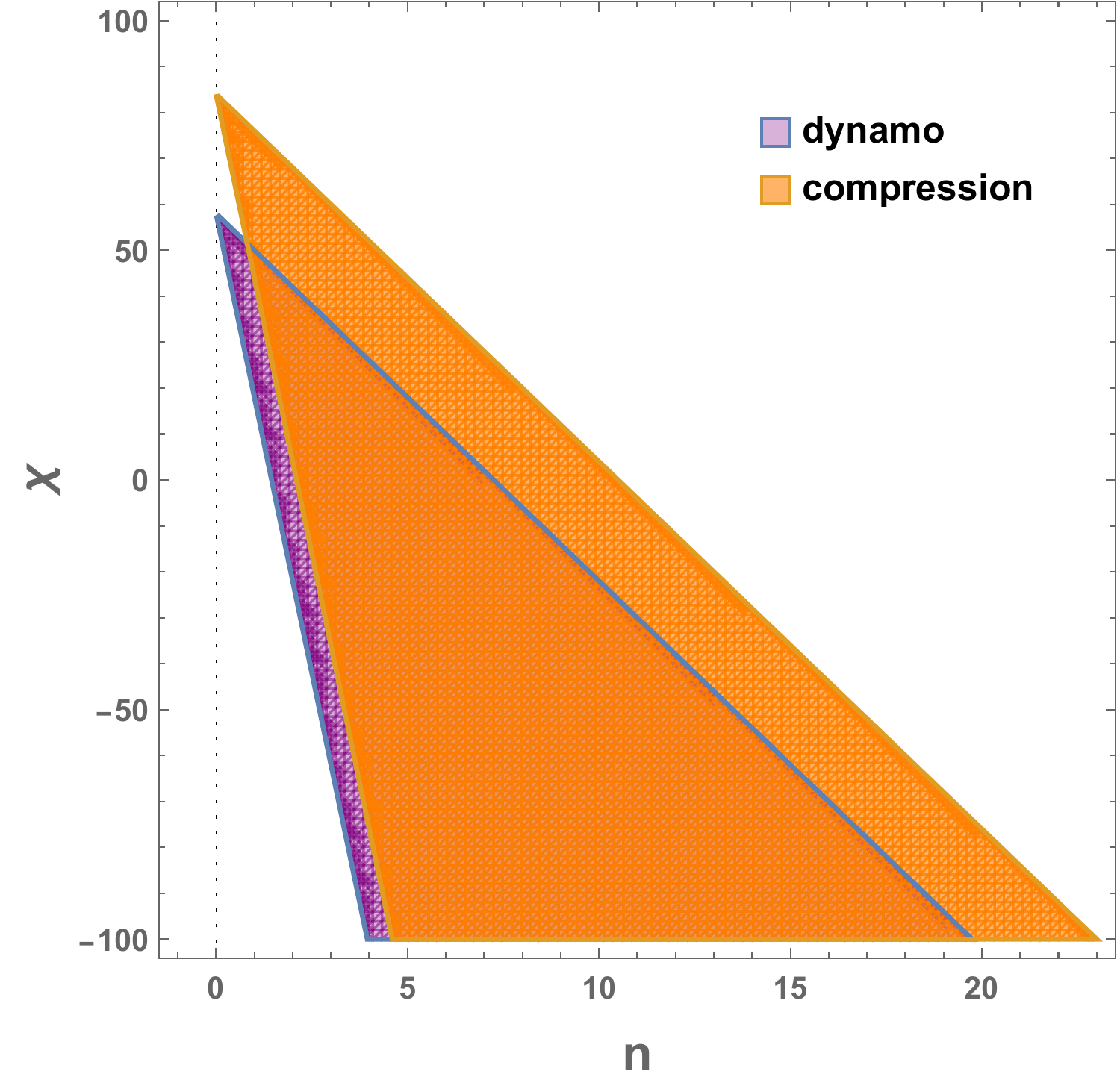}
\caption{\label{fig:2} Compatibility regions for the model power $n$ and the inflation exponent $\chi$ assuming a dynamo mechanism (purple) or a compression effect (orange) on the galactic seed fields. Notice that, at a fixed power $n$,  wide intervals are allowed to $\chi$, which becomes less restrictive as $n$ increases  }
\end{figure}

We have to remember that, in the first case, a (galactic) dynamo mechanism is invoked, while in the second case the amplification is achieved during the collapse of the protogalactic cloud. The above inequalities are shown in Fig. \ref{fig:2}. Notice that, at a fixed power $n$,  wide intervals are allowed to $\chi$, which  become less restrictive as $n$ increases. In particular, when $n=3$, one recovers results in \cite{Bertolami:2022hjk}. The fact to obtain similar results in metric and  teleparallel theories  is not surprising: thanks to the slow-roll approximation, the boundary term $B$ plays the same role of the Ricci scalar $R$, thus gaining its own importance even in the absence of the torsion scalar $T$. By the way, considering a non-minimal coupling function like $f_{2}(T,B)=-\lambda T B^{n}$, after similar calculations, we arrive at the same inequalities with the only change $2n \mapsto 2n+2$, thus getting even less tight constraints.

\section{Discussion and Conclusions}\label{sec6}
Today, it is well known that stars or compact objects (like neutron stars) are capable of generating very strong local magnetic fields (up to $10^{14}$ G). The increasingly evident presence of galactic and especially intergalactic magnetic fields, however, still have no explanation, being one of the long standing puzzle of astrophysics and cosmology.  While in the case of aged galaxies, one could explain them by invoking a dynamo \cite{Widrow_2002} or a compression mechanism \cite{Turner:1987bw}, the presence of fields also in protogalaxy and in the intergalactic medium, suggest a  cosmological rather than local origin of them. A recent observational analysis \cite{DiGennaro:2020zkz} found high radio luminosities in high redshift clusters, suggesting that magnetic field amplification happens  during the first phases of cluster formation. Indeed, a direct link between PMFs (as constrained by the CMB anisotropy power spectra) and the nowadays cosmic
magnetism  was also confirmed by simulations \cite{Vazza:2020phq}. This status of art suggests some today  unknown physical process for generating such large scale fields. One possibility is that they are relics from the early Universe, with a subsequent amplification in a pregalactic era. Inflation remains  the only epoch capable of  producing super-horizon correlations, in order to generate such large scale fields. These long-wavelength modes, in particular, may have been generated by quantum fluctuations which have grown  during inflation and  the reheating eras,  in a similar way to that which led to  density fluctuations  and thus to the large scale structure of the Universe. Their effect on gravitational waves  and on the CMB has been studied in \cite{GW-2021al} and \cite{Kunze:2013kza}, respectively. The standard GR predicts an adiabatic decay for this primordial field: since the Universe is believed to have been a good conductor for much of its  post-inflationary history, any  cosmological
magnetic field, from the end of inflation onwards,  will preserve the  flux, i.e. $a^2 \textbf{B} \sim const $, and then $\textbf{B} \sim 1/a^2$ (adiabatic decay). The same fate occurs during inflation, as is evident by solving Maxwell's equations. Since $a$ grows exponentially during inflation, this means that GR  cannot explain their survival. Hence the need to move towards extended gravity theories.

In this paper, we studied the amplification of PMFs in the context of  extended teleparallel theories of gravity, i.e. $f(T,B)$ gravity, which recently revealed extremely useful to fix several cosmological issues.  

In the first part, we computed and solved the exact cosmological equations in a spatially flat FRW metric, distinguishing between inflationary and reheating eras. We used two different models, namely $f(T,B)=-T+\lambda B^{n}$  and $f(T,B)=-\lambda T B^{n}$. Here, we found important deviations from GR, both in inflation and in the reheating era. In the inflationary phase,  for the first model and in the strong field  regime $\lambda/M_{pl}^{2(1-n)} \gg 1$, we showed that the time power $\alpha$ of the scale factor $a(\eta)==1/(-c \eta)^{\alpha}$ is such that $\alpha>1$ $\forall n>1$, implying a faster inflation. For the second model, instead, a distinction between the various regimes of the coupling constant $\lambda$ is not necessary, but the solution for a generic equation of state  parameter $w$ is very involved. For this model, we found $\alpha \geq 1 $ $\forall n >0$. In the reheating phase, a solution for $\alpha$ exists only in the range $7/16 \leq n<2$ (see Fig. \ref{fig:1} (a)) for the first model, while  an analytical solution is not possible for the second model (see Table~\ref{tab1}).

In the second part, we adopt a non-minimal gravity-photon coupling in order to generate primordial magnetic fields with a non-adiabatic behaviour, exploiting the breaking of conformal symmetry (which naturally arises from such non-minimal couplings) and the results from the first part.  These couplings are well motivated, since according to the QED on curved space-time, one-loop vacuum-polarization effects can lead to  non-minimal gravitational couplings between the curvature and the electromagnetic field \cite{Drummond:1979pp}.  The electromagnetic sector $F_{\mu \nu}F^{\mu \nu}$ was coupled to the  scalar $R$ and tensor $R^{\mu \nu \rho \sigma}$ curvatures in \cite{Turner:1987bw,Lambiase:2008zz,deAndrade:2013fga}, curvature power-law models $R^{n}$ in \cite{Garretson:1992vt,Mazzitelli:1995mp,Lambiase:2008zz,Bertolami:2022hjk}, torsion power-law models $T^{n}$ in \cite{Bamba:2013rra}. It was proposed in \cite{Pavlovic:2018idi} that the signatures of such non-minimal couplings could be in principle observed  (or constrained) by investigating  the magnetic fields around the event horizons of black holes and that the same effect could be exploited for constraining the size of primordial black holes.  

After obtaining the corresponding Maxwell equations, we studied the evolution of  magnetic field $\mathbf{B}$ in the two epochs. We assumed a negligible non-minimal coupling during inflation, so that an amplification effect is realized only during the reheating era, where, for the first $f(T,B)$ model (now coupled to the electromagnetic sector), we found that amplification is always possible ($\gamma>1$), as showed in Fig. \ref{fig:1} (b), and any adiabatic decrease is avoided. The amplification effect is even more evident in the second model, especially for $n>1/2$. Interestingly, we found that imposing a zero electric field, automatically leads to the super-horizon approximation, $k \eta \leq 1$, while the inverse is guaranteed (approximately) only for inflation phase. 

We finally considered results in \cite{Bertolami:2022hjk}, in order to estimate amplification of the magnetic field assuming its manifestation during \textit{inflation}, i.e. turning on the non-minimal coupling in this epoch, rather than in the \textit{reheating} one. In order to explain the present value of the rate between the magnetic energy density and cosmic microwave energy density, $r=\rho_{B}/ \rho_{\gamma} \simeq 1$, we found that, during inflation, it should have been $r_{IN}=10^{\chi}$, with $\chi \geq 58 - 40 n$, if a (galactic) dynamo mechanism is invoked, and $\chi \geq 84 - 40 n$ if the amplification is achieved during the collapse of the protogalactic cloud. Contextually, we found that thanks to the slow-roll approximation, the boundary term $B$ plays the same role of the Ricci scalar $R$, thus gaining its own importance even in the absence of the torsion scalar $T$. 

Considering an electromagnetic tensor in a space-time with torsion is  a very delicate task. Since the connection symbols are no longer symmetric, the standard definition of $F_{\mu \nu}$ is no more compatible with the $U(1)$ gauge invariance of QED, thus requiring a change in the minimal prescription $\partial_{\mu} \rightarrow \nabla_{\mu}$ (at the action level) to switch  from flat to curved space-times (see \cite{Fresneda:2014kua} for a  discussion and references therein). This aspect is not often emphasized when dealing with electromagnetism in cosmological  backgrounds with torsion and neglecting it implicitly  means that photons do not react to torsion. Although this aspect is not considered in this work, it certainly constitutes an important issue to be taken into account in a general discussion of magnetic fields in presence of torsion. We plan to study this topic in a forthcoming paper.

\begin{acknowledgments}

The authors acknowledge the  support by the  Istituto Nazionale di Fisica Nucleare (INFN) {\it Iniziativa Specifica} QGSKY. 
\end{acknowledgments}

\appendix

\section{The  torsion scalar }

In this appendix, we report in details the calculations needed to derive the second equality in Eq.~(\ref{eq17}), using the definition of the torsion tensor.  
Let us start with  the definition (\ref{eq6}) of the scalar torsion
\begin{equation}\label{eqA1}
T=\dfrac{1}{4}T\indices{^\rho ^\mu ^\nu}T\indices{_\rho _\mu _\nu} + \dfrac{1}{2}T\indices{^\rho ^\mu ^\nu}T\indices{_\nu _\mu _\rho} -T\indices{^\rho _\mu _\rho}T\indices{^\nu ^\mu _\nu} .  
\end{equation}
The easier term to compute is 
\begin{equation*}
    T\indices{^\rho _\mu _\nu}=h\indices{_a ^\rho}\left(\partial_{\mu} h\indices{^a _\nu}-\partial_{\nu} h\indices{^a _\mu}\right)
 \end{equation*}   
 which, in the conformal metric Eq.~(\ref{eq14}),  becomes
 \begin{equation*}
   T\indices{^\rho _\mu _\nu}=  \frac{a'}{a}\Big(\delta_{0}^{\rho} \delta_{\nu}^{0} \delta_{\mu}^{0} -\delta_{0}^{\rho} \delta_{\mu}^{0} \delta_{\nu}^{0} \Big)+\frac{1}{a}\Big[\delta_{i}^{p} \partial_{\mu}\left(a \delta_{\nu}^{i}\right)-\delta_{i}^{\rho} \partial_{\nu}(a \delta_{\mu}^{i})\Big]\,,
\end{equation*}   
that is
\begin{equation*}
 T\indices{^\rho _\mu _\nu}=   \mathcal{H}\left( \delta_{i}^{\rho}\delta_{\nu}^{i}\delta_{\mu}^{0} - \delta_{i}^{\rho}\delta_{\mu}^{i}\delta_{\nu}^{0} \right) a
\end{equation*}
where we used the definition of the Hubble constant in the conformal time,  $\mathcal{H}= a'/a^{2}$ . 
The next component we need is
\[\arraycolsep=1.4pt\def\arraystretch{2.2}
\begin{array}{ccc}
 T \indices{_\rho _\mu _\nu}= g_{\lambda\rho}T\indices{^\lambda _\mu _\nu} = \\
 g_{0\rho}T\indices{^0 _\mu _\nu} + g_{j \rho}T\indices{^j _\mu _\nu}= - a^{3} \mathcal{H} \cdot \\ 
\Big[ \delta_{\rho}^{1}\Big(\delta_{\nu}^{1} \delta_{\mu}^{0}-\delta_{\mu}^{1} \delta_{\nu}^{0}\Big) + \delta_{\rho}^{2}\Big(\delta_{\nu}^{2} \delta_{\mu}^{0}-\delta_{\mu}^{2} \delta_{\nu}^{0}\Big)+ \delta_{\rho}^{3}\Big(\delta_{\nu}^{3} \delta_{\mu}^{0}-\delta_{\mu}^{3} \delta_{\nu}^{0}\Big) \Big].
 \end{array}
 \]
Similarly, one finds 
\[\arraycolsep=1.4pt\def\arraystretch{2.2}
\begin{array}{ccc}
    T \indices{^\rho ^\lambda _\nu}=g^{\lambda\mu}T\indices{^\rho _\mu _\nu} = \\
    \dfrac{\mathcal{H}}{a} \Big( \delta_{0}^{\lambda}\delta_{i}^{\rho}\delta_{\nu}^{i}  + \delta_{1}^{\lambda}\delta_{1}^{\rho}\delta_{\nu}^{0} + \delta_{2}^{\lambda}\delta_{2}^{\rho}\delta_{\nu}^{0} + \delta_{3}^{\lambda}\delta_{3}^{\rho}\delta_{\nu}^{0}\Big).
 \end{array}
 \]
Finally,
\[\arraycolsep=1.4pt\def\arraystretch{2.2}
\begin{array}{ccc}
    T \indices{^\rho ^\lambda ^\sigma}=g^{\sigma\nu}T\indices{^\rho ^\lambda _\nu} = \\
    \dfrac{\mathcal{H}}{a^{3}} \Big[ \delta_{0}^{\sigma}\delta_{1}^{\lambda}\delta_{1}^{\rho}  + \delta_{0}^{\sigma}\delta_{2}^{\lambda}\delta_{2}^{\rho} + \delta_{0}^{\sigma}\delta_{3}^{\lambda}\delta_{3}^{\rho} - \delta_{0}^{\lambda} \Big( \delta_{1}^{\sigma}\delta_{1}^{\rho} + \delta_{2}^{\sigma}\delta_{2}^{\rho} + \delta_{3}^{\sigma}\delta_{3}^{\rho}\Big)\Big].
 \end{array}
 \]
We can now  compute all the terms in Eq.~(\ref{eqA1}). The first term is
\[\arraycolsep=1.4pt\def\arraystretch{2.2}
\begin{array}{ccc}
    T \indices{^\rho ^\mu ^\nu} T \indices{_\rho _\mu _\nu}= \\ -\mathcal{H}^{2}\Big[\delta_{0}^{\nu}\Big( \delta_{1}^{\mu} \delta_{1}^{\rho}+ \delta_{2}^{\mu} \delta_{2}^{\rho}+ \delta_{3}^{\mu} \delta_{3}^{\rho}\Big)- \delta_{0}^{\mu} \Big( \delta_{1}^{\nu}  \delta_{1}^{\rho}-  \delta_{2}^{\nu}  \delta_{2}^{\rho}-\delta_{3}^{\nu}  \delta_{3}^{\rho}\Big)\Big]  \\
 \Big[\delta_{\mu}^{0} \Big(\delta_{\rho}^{1} \delta_{\nu}^{1}   
 +\delta_{\rho}^{2}\delta_{\nu}^{2}
 +\delta_{\rho}^{3} \delta_{\nu}^{3} \Big)
 -\delta_{\nu}^{0}\Big(\delta_{\rho}^{1} \delta_{\mu}^{1} 
 +\delta_{\rho}^{2} \delta_{\mu}^{2} 
 +\delta_{\rho}^{3} \delta_{\mu}^{3} \Big) \Big]=6\mathcal{H}^{2}.
 \end{array}
 \]
Similarly, the second term  reduces to
\[\arraycolsep=1.4pt\def\arraystretch{2.2}
\begin{array}{ccc}
    T \indices{^\rho _\mu _\rho} T \indices{^\nu ^\mu _\nu}= \\ 
    -\mathcal{H}^{2}\Big[\delta_{0}^{\nu}\Big( \delta_{1}^{\mu} \delta_{1}^{\rho}+ \delta_{2}^{\mu} \delta_{2}^{\rho}+ \delta_{3}^{\mu} \delta_{3}^{\rho}\Big)- \delta_{0}^{\mu} \Big( \delta_{1}^{\nu}  \delta_{1}^{\rho}-  \delta_{2}^{\nu}  \delta_{2}^{\rho}-\delta_{3}^{\nu}  \delta_{3}^{\rho}\Big)\Big]  \\
 \Big[\delta_{\rho}^{0} \Big(\delta_{\rho}^{1} \delta_{\nu}^{1}   
 +\delta_{\rho}^{2}\delta_{\nu}^{2}
 +\delta_{\rho}^{3} \delta_{\nu}^{3} \Big)
 -\delta_{\rho}^{0}\Big(\delta_{\nu}^{1} \delta_{\mu}^{1} 
 +\delta_{\nu}^{2} \delta_{\mu}^{2} 
 +\delta_{\nu}^{3} \delta_{\mu}^{3} \Big) \Big]=3\mathcal{H}^{2},
 \end{array}
 \]
while the last term is 
\[
 T \indices{^\rho _\mu _\rho} T \indices{^\nu ^\mu _\nu}=  -\mathcal{H}^{2}\Big[\delta_{i}^{\rho} \delta_{\rho}^{i} \delta_{0}^{\mu}\Big]  
 \Big[\delta^{\mu}_{0} \delta_{i}^{\nu} \delta_{\nu}^{i}\Big]=9\mathcal{H}^{2}
\]
 
Putting all together in Eq.~(\ref{eqA1}), one arrives to 
\begin{equation}
T=-6\mathcal{H}^{2}. 
\end{equation}  Notice that in the cosmological time, it is $T=-6 H^{2}$, and the two relations are clearly related through the variable change $t \mapsto \eta$. 

\bibliography{biblio2}

\end{document}